# On the Rabinowicz like criterion of formation of wear particles in a system with a soft surface layer

Valentin L. Popov

*Technische Universität Berlin,*
*Department of System Dynamics and the Physics of Friction*
*10623 Berlin, Germany*
*e-mail: v.popov@tu-berlin.de*

In 1958, Ernest Rabinowicz suggested a simple criterion distinguishing the regimes of plastic smoothing and formation of wear particles in a contact of homogeneous sliding bodies. However, he did not consider any detailed mechanism of either plastic smoothing or debris formation. In a recent paper in Nature Communications, Molinari et al. have confirmed the criterion using explicit mesoparticle simulation. The work of Molinari's group provides a strong support to the general concept of Rabinowicz which is based on the consideration of competition of plastic deformation and fracture. It is interesting to apply this concept to more general configurations than those considered by Rabinowicz and Molinari's group. An important case is a system having a very soft surface layer. In the present paper, the analysis similar to that of Rabinowicz is applied to materials with soft layer. It is shown that in this case, too, both cases of plastic deformation and debris formation may occur. It would be interesting to verify this prediction by explicit meso-particle simulations similar to those of Molinari's group.

## I. Introduction

In 1958, E. Rabinowicz wrote a short article in Wear [1] containing only a couple of simple equations, in which he put forward the hypothesis of a mechanism determining the size of wear particles. If two asperities collide and form a welded bridge, as suggested by Bowden and Tabor [2], they are first plastically deformed, until the maximum stress is achieved, which is on the order of magnitude of the hardness of the material. In this state, there is some elastic energy stored in the material which is proportional to the third power of contact size. This energy can relax by forming a wear particle, but only if the stored elastic energy exceeds the energy needed to create new free surfaces, the latter being proportional to the square of the junction size. This leads to the conclusion that only particles larger than some critical particle size can be created. Rabinowicz did not explain the mechanism of transition from wear less sliding to particle formation. Decades later, the group of Molinari managed to show by direct quasi-molecular simulations that the transition does exist [3] and shed light to the detailed process of plastic smoothing. Based on this work, Popov [4] suggested a very simple model of competition of plasticity and fracture, also leading to the Rabinowicz criterion but simplifying the process to the minimum possible elementary process and thus illustrating clearly the details of the process. According to this model, some plastic deformation will always precede a possible cracking process, and it is this process of plastic flow which limits the possible stress concentration needed for the opening and propagating of a crack.

Rabinowicz has formulated his criterion for homogeneous media. However, many tribological systems have a pronounced layered structure – either artificially designed or developed during the tribological loading. In the present paper we repeat the arguments of Rabinowicz for such layered system and find the conditions for plastic smoothing and particle detachment in this case.

## II. Model of a layered system

Consider an elastic medium with elastic shear modulus $G_0$ covered with a soft elastoplastic layer of thickness $h$ having shear modulus $G_c$ and the tangential yield stress $\sigma_c$. This layer can be deposited to the surface artificially or it can appear naturally through mechanically induced chemical reactions of the base material with surrounding substances (lubricant, counter-body, air and so on) [5],[6]. Assume that due to normal loading and tangential sliding a junction of the diameter $D$ is formed. The components of the stress tensor in the near surroundings of the junction will be on the order of magnitude of $\sigma_c$. If the elastic energy stored in the system is not enough for creating new surfaces having the area of the order of $D^2$ than the only possible process will be plastic smoothing as illustrated in detail in the paper [3]. In the opposite case, the elastic energy can be relaxed by detaching of a wear particle. In the case of debris formation, two limiting cases are possible:

*1. The detachment occurs in the base material.*

In this case, we basically can repeat the line of arguments of Rabinowicz. The stored elastic energy has the order of $\frac{\sigma_c^2}{2G_0}D^3$ and the surface energy needed for formation of a wear particle of the order of $\gamma_0 D^2$ where $\gamma_0$ is the work of separation of the base material. The formation of wear particle is possible if $\frac{\sigma_c^2}{2G_0}D^3 > \gamma_0 D^2$ or

$$D > \frac{2G_0 \gamma_0}{\sigma_c^2}, \qquad (1)$$

that coincides with the Rabinowicz criterion (see [3],[4]). Note that the only difference from the classical criterion is that the elastic modulus and energy of separation are those of the base material while the critical flow stress is that of the surface layer.

*2. The detachment occurs inside the surface layer.*

In this case, the elastic energy which is released due to particle detachment is on the order of $\frac{\sigma_c^2}{2G_c}D^2 h$ and the energy needed for detachment $\gamma_c D^2$. The formation of particles is thus possible if $\frac{\sigma_c^2}{2G_c}D^2 h > \gamma_c D^2$ or

$$h > h_c = \frac{2G_c \gamma_c}{\sigma_c^2}. \qquad (2)$$

In this criterion the fulfilment of criterion does not depend on the diameter of junction but depends solely on the *thickness* of the layer. *If thickness of the layer is smaller than the critical one, $h_c$, than formation of particles is not possible independently of the size of junctions.* Note that in this case only the properties of the softer surface layer do play a role.

### III. Criteria for formation of "flat" and "spherical" wear particles

Let us consider in detail the transition between the cases 1 and 2 considered in the previous Section. Again consider a junction with some particular diameter $D$. The following cases are possible:

1. $D > \frac{2G_0 \gamma_0}{\sigma_c^2}$ but $h < \frac{2G_c \gamma_c}{\sigma_c^2}$. In this case, formation of the in-layer particles is not possible but formation of the base-material particles is possible. This is the classical "Rabinowicz case".

2. $D < \frac{2G_0 \gamma_0}{\sigma_c^2}$ but $h > \frac{2G_c \gamma_c}{\sigma_c^2}$. (This case is possible if the elastic modulus of the surface layer is sufficiently smaller than that of the base material). In this case, the formation of "bulk" particles is not possible but the in-layer flat wear particles can be formed.

3. In the general case, one could suggest the following generalized estimation. Assume that we have a junction of diameter $D$ and detached is a particle of diameter $D$ and thickness $H$. Then, the elastic energy stored in the system is on the order of

$$U_{el} = \begin{cases} \dfrac{D^2\sigma_c^2}{2}\left(\dfrac{h}{G_c}+\dfrac{H-h}{G_0}\right), & \text{for } H>h \\ \dfrac{D^2\sigma_c^2}{2}\dfrac{H}{G_c}, & \text{for } H<h \end{cases} \tag{3}$$

The energy needed for formation of the above particle is on the order of

$$U_{surf} = \begin{cases} \gamma_0 D^2, & \text{for } H>h \\ \gamma_c D^2, & \text{for } H<h \end{cases} \tag{4}$$

The formation of particles is possible if

$$\begin{cases} \dfrac{D^2\sigma_c^2}{2}\left(\dfrac{h}{G_c}+\dfrac{H-h}{G_0}\right) > \gamma_0 D^2, & \text{for } H>h \\ \dfrac{D^2\sigma_c^2}{2}\dfrac{H}{G_c} > \gamma_c D^2, & \text{for } H<h \end{cases} \tag{5}$$

or

$$\begin{cases} H > \dfrac{2G_0\gamma_0}{\sigma_c^2} - h\left(\dfrac{G_0}{G_c}-1\right), & \text{for } H>h \\ H > \dfrac{2G_c\gamma_c}{\sigma_c^2}, & \text{for } H<h \end{cases} \tag{6}$$

Let us display these relations graphically on the plane $(H,h)$,

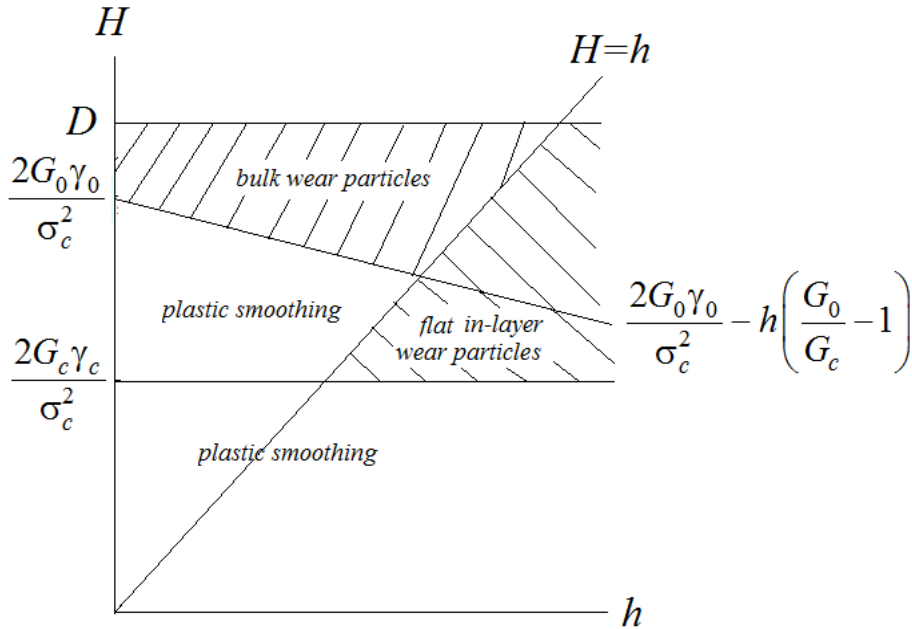

**Figure 1.** Schematic representation of conditions given by Eq. (6).

A completely "wear-less" sliding will occur if the following two conditions are fulfilled:

$$h < \frac{2G_c\gamma_c}{\sigma_c^2} \tag{7}$$

and

$$D < \frac{2}{\sigma_c^2}\left[G_0\gamma_0 - G_0\gamma_c + G_c\gamma_c\right]. \tag{8}$$

Due to softness of the surface layer, the critical junction size can be made large enough so that the only condition which has to be observed would be that given by Eq. (7). This explains the principle of "positive hardness gradient" as condition for wear less sliding, which was formulated by Kragelsky [7].

Of course, even the process of plastic smoothing will lead to effective "wear" due to "squeezing out" of the surface layer. However, it was shown in [5] (see also [8], §17.5) that in this case the effective wear rate is proportional to the square of the ratio of the layer thickness to the linear size $L$ of the frictional contact zone. The wear coefficient will thus be on the order of

$$k_{adh} \propto \left(\frac{h}{L}\right)^2 \quad (9)$$

and can assume extremely small values. Note that the smaller is the thickness of the surface layer the smaller is the wear coefficient.

## IV. Conclusion

In the present note we applied the logic used by Rabinowicz in his famous derivation of the critical junction size for production of wear particles to an elastic material covered by a soft elasto-plastic layer. We have shown that in this case, the regime is possible when the formation of wear particles is impossible – neither of bulk particles not of in-layer flat particles. This regime is realized by thin enough surface layers. This result explains the known rule put forward by Kragelsky and confirmed by Gerve, Kehrwald and Scherge that the least wear is observed if a very thin soft layer is formed during the running-in process of the system.